\documentclass[lettersize,journal]{IEEEtran}
\usepackage{xspace}
\usepackage{cite}
\usepackage{amsmath,amssymb,amsfonts}
\usepackage{algpseudocode}
\usepackage[ruled]{algorithm}
\usepackage{graphicx}
\usepackage{textcomp}
\usepackage{booktabs}
\usepackage{diagbox}
\usepackage{caption}
\usepackage{listings, xcolor, soul} 
\usepackage{mathrsfs}
\usepackage{multirow}
\usepackage{ifthen}
\usepackage{color, colortbl}
\def\BibTeX{{\rm B\kern-.05em{\sc i\kern-.025em b}\kern-.08em
		T\kern-.1667em\lower.7ex\hbox{E}\kern-.125emX}}

\AtBeginDocument{%
	\providecommand\BibTeX{{%
			\normalfont B\kern-0.5em{\scshape i\kern-0.25em b}\kern-0.8em\TeX}}}
\usepackage{inconsolata}
\usepackage{enumitem}
\usepackage{hyperref}
\newtheorem{definition}{Definition}

\let\origthelstnumber\thelstnumber
\makeatletter
\newcommand*\Suppressnumber{%
	\lst@AddToHook{OnNewLine}{%
		\let\thelstnumber\relax%
		\advance\c@lstnumber-\@ne\relax%
	}%
}

\newcommand*\Reactivatenumber[1]{%
	\setcounter{lstnumber}{\numexpr#1-1\relax}
	\lst@AddToHook{OnNewLine}{%
		\let\thelstnumber\origthelstnumber%
		\refstepcounter{lstnumber}
	}%
}

\makeatother

\definecolor{Gray}{gray}{0.6}
\definecolor{LightGray}{gray}{0.9}
\definecolor{bananayellow}{rgb}{1.0, 0.88, 0.5}

\newcommand{\toolname}{\textsc{Selertion}\xspace}

\newboolean{showcomments}
\setboolean{showcomments}{true}
\ifthenelse{\boolean{showcomments}}
{
	\definecolor{myyellow}{RGB}{255, 228, 26}
	\definecolor{myblue}{RGB}{50, 50, 220}
	\newcommand{\nb}[2]{
		{\sf
                \color{myblue}{#2}
		}%
	}
}
{
	\newcommand{\nb}[2]{}
}

\newcommand{\header}[1]{\par\smallskip\noindent\textbf{#1}}
\begin{document}
	\title{Fine-Grained Assertion-Based Test Selection}
	\author{Sijia Gu, Ali Mesbah
		\thanks{S. Gu is with Department of Electrical and Computer Engineering, University of British Columbia, Vancouver, Canada.}
		\thanks{A. Mesbah is with Department of Electrical and Computer Engineering, University of British Columbia, Vancouver, Canada}
	}
	\maketitle
	
	\begin{abstract}
		For large software applications, running the whole test suite after each code change is time- and resource-intensive. Regression test selection techniques aim at reducing test execution time by selecting only the tests that are affected by code changes. However, existing techniques select test entities at coarse granularity levels such as test class, which causes imprecise test selection and executing unaffected tests. We propose a novel approach that increases the selection precision by analyzing test code at statement level and treating test assertions as the unit for selection. We implement our fine-grained test selection approach in a tool called \toolname and evaluate it by comparing against two state-of-the-art test selection techniques using 11 open-source subjects. Our results show that \toolname increases selection precision for all the subjects.  Our test selection reduces, on average, 63\% of the overall test time, making regression testing 7--38\% faster than the other techniques. Our results also indicate that subjects with longer test execution time benefit more by our fine-grained selection technique. 
	\end{abstract}
	
	
	\section{Introduction}
	Regression testing has been a vital process to ensure no existing functionality is broken when software evolves. However, even for large companies with allocated enormous resources to testing~\cite{engstrom2010qualitative, elbaum2014techniques, memon2017taming}, it can be costly to rerun the whole test suite every time for the code changes. For example, Elbaum et al. \cite{elbaum2000prioritizing} report that it costs seven weeks to execute the entire test suite for one of their industrial partner's products. In addition to companies that conduct tests in continuous integration environments, individual developers often find themselves in a situation where they need to run regression tests multiple times a day before committing their changes to the codebase~\cite{gligoric2014empirical}. 
	
	As a result, test suite optimization techniques that reduce test runtime such as regression test selection, minimization, and prioritization have become an inalienable part of software research and development~\cite{engstrom2010qualitative, kiran2019comprehensive, yoo2012regression}. Regression test selection (RTS)~\cite{kazmi2017effective, gligoric2015practical} in particular aims to run only the tests that are affected by code changes. 
	RTS techniques aim to be (1) safe, by selecting all tests affected by code changes, and (2) precise, by selecting fewer tests. 
	
	Current techniques~\cite{ren2004chianti, zhang2011localizing, gligoric2015practical, zhang2018hybrid} analyze code at different levels of granularity. Coarse-grained RTS techniques that analyze code at file-level~\cite{gligoric2015practical} incur less analysis overhead but select more tests, i.e. they are less precise  than finer-grained RTS techniques that analyze code at basic block or method level~\cite{ren2004chianti, zhang2011localizing}. However, existing RTS techniques typically focus on the granularity of production code (code under test), while select test entities at a relatively coarse granularity such as the test-class level. This can result in safe but imprecise test selection and, consequently, longer test execution times. The root cause of this issue lies in the fact that, in practice, only a portion of the test code within the selected test classes is impacted by code changes. For instance, when an atomic production method \texttt{negate} is changed in the \texttt{Complex} class from Apache Commons Math~\cite{commonsMath}, the coarse-grained RTS techniques~\cite{zhang2018hybrid, gligoric2015practical} select the whole test class~\texttt{ComplexTest} with 138 test cases. However, only seven out of the 138 test cases are truly affected by the changed method. Furthermore, empirical studies~\cite{zhang2015assertions} have shown that it is a common practice for a single test case to contain multiple assertions. Therefore, only specific assertions within a test case might be influenced by code changes, leaving others unaffected.
	
	Our insight is that a finer-grained test code analysis can help to increase selection precision and accelerate test execution. As such, we focus on test methods and test assertions inside test methods, which are known to be correlated with a test suite's fault detection ability~\cite{zhang2015assertions}. We propose a novel test selection technique, called \toolname, which analyzes test entities at the fine-grained test assertion level for objected-oriented Java programs. We treat each assertion inside a test method as the smallest unit of interest and slice the test method into executable fragments containing the targeted assertion and statements associated with it for selection. 
	In comparison to coarse-grained RTS techniques, \toolname  selects fewer tests more precisely, ultimately leading to a more efficient test execution process.
	
	
	Our work offers the following significant contributions:
	\begin{itemize}
		\item The first regression test selection technique that operates at the finest granularity, treating test assertions within test methods as the fundamental units for selection.
		\item A method for statically slicing test methods into executable fragments for testing each assertion.
		\item An implementation of our approach in a tool called \toolname, which supports Java projects with JUnit4 tests.
		\item An empirical evaluation of \toolname performed on 11 open-source projects and comparison with two state-of-the-art RTS tools. 
	\end{itemize}
	
	Our evaluation results reveal that \toolname selects 15.8\% of tests, representing a precision improvement of 43\%--53\% compared to existing state-of-the-art test selection techniques. Furthermore, \toolname significantly enhances efficiency by reducing, on average, 63\% of the overall test execution time, resulting in a speed improvement of 7--38\%.
	
	
	\section{Motivating Example}\label{example}
	
	To illustrate our insight, we use real test cases from the \texttt{ComplexTest} class of Apache Commons Math \cite{commonsMath} as a motivating example. Figure \ref{fig1} shows two test methods \texttt{testExp} and \texttt{testNegate} that both have multiple assertions (\textit{i.e.,} Lines \ref{line:5}, \ref{line:6}, \ref{line:8} in \texttt{testExp} and Lines \ref{line:15}, \ref{line:16} in \texttt{testNegate}). When the production method \texttt{Complex.negate} is changed, all the 138 test cases in \texttt{ComplexTest} class will be selected if the selection is at test-class level. However, only seven (including the two shown in Figure \ref{fig1} ) out of the 138 test cases are truly affected by the changed production method. Moreover, when focusing on the assertions in each test case, the first two assertions in \texttt{testExp} at Lines \ref{line:5} and \ref{line:6} neither check \texttt{Complex.negate}'s behavior nor have a dependency on it. Hence, it is not necessary to execute these two unaffected assertions. Meanwhile, if the two assertions in \texttt{testNegate} are assumed to fail because of the changes in \texttt{Complex.negate}, the execution will stop at the first failed assertion at Line \ref{line:15}, and the developer may overlook the second failed assertion at Line \ref{line:16} when fixing the bug in \texttt{Complex.negate}.  More time will be consumed by executing the test suite again to reveal the second failure at Line \ref{line:16}.
	
	The fine-grained test selection approach we propose is to divide a test method into smaller code fragments called \textit{assertion slices} as the smallest units for selection. Each \textit{assertion slice} includes the test assertion and previous statements associated with it. The assertion slices for a test method can be obtained by performing static intra-procedural backward slicing \cite{weiser1984program} on the variables of each assertion as the selected points of interest. Figure \ref{fig2} presents the assertion slices for the two test methods \texttt{testExp} and \texttt{testNegate} in Figure \ref{fig1} as an example. The grayed-out slices in Figure \ref{fig2} are the ones that are selected when there are changes in the production method \texttt{Complex.negate}. Our fine-grained assertion-level selection technique selects only the assertions depending on the code changes and guarantees that all the selected assertions will be checked. Next we will present a more detailed description of our approach.
	\begin{figure}
		\begin{lstlisting}
		@Test
		public void testExp() {
		Complex z = new Complex(3, 4);(*@\label{line:3}@*)
		Complex expected = new Complex(-13.12878, -15.20078);
		(*@\hl{TestUtils.assertEquals(expected, z.exp(), 1.0e-5);} @*)(*@\label{line:5}@*)
		(*@\hl{TestUtils.assertEquals(Complex.ONE,Complex.ZERO.exp(), 10e-12);}@*)(*@\label{line:6}@*)
		Complex iPi = Complex.I.multiply(new Complex(pi,0));(*@\label{line:7}@*)
		(*@\hl{TestUtils.assertEquals(Complex.ONE.negate(),iPi.exp(), 10e-12);}@*)(*@\label{line:8}@*)
		}
		
		@Test
		public void testNegate() {
		Complex x = new Complex(3.0, 4.0);
		Complex z = x.negate();
		(*@\hl{Assert.assertEquals(-3.0, z.getReal(), 1.0e-5);}@*)(*@\label{line:15}@*)
		(*@\hl{Assert.assertEquals(-4.0, z.getImaginary(), 1.0e-5);}@*)(*@\label{line:16}@*)
		}
		// 2 of 138 test cases from ComplexTest.java
		\end{lstlisting}
		\caption{Test cases from Apache Commons Math. Assertion statements are highlighted.}
		\label{fig1}
	\end{figure}
	
	\begin{figure}
		\begin{lstlisting}[firstnumber=3]
		Complex z = new Complex(3, 4);
		Complex expected = new Complex(-13.12878, -15.20078);
		(*@\hl{TestUtils.assertEquals(expected, z.exp(), 1.0e-5);} @*)
		\end{lstlisting}
		\begin{lstlisting}[firstnumber=6]
		(*@\hl{TestUtils.assertEquals(Complex.ONE,Complex.ZERO.exp(), 10e-12);}@*)
		\end{lstlisting}
		\begin{lstlisting}[firstnumber=7, backgroundcolor=\color{gray!30}]
		Complex iPi = Complex.I.multiply(new Complex(pi,0));
		(*@\hl{TestUtils.assertEquals(Complex.ONE.negate(),iPi.exp(), 10e-12);}@*)
		\end{lstlisting}
		
		\begin{lstlisting}[firstnumber=13, backgroundcolor=\color{gray!30}]
		Complex x = new Complex(3.0, 4.0);
		Complex z = x.negate();
		(*@\hl{Assert.assertEquals(-3.0, z.getReal(), 1.0e-5);}@*)
		\end{lstlisting}
		
		\begin{lstlisting}[firstnumber=13, backgroundcolor=\color{gray!30}]
		Complex x = new Complex(3.0, 4.0);
		Complex z = x.negate();(*@\Reactivatenumber{15}@*)
		(*@\hl{Assert.assertEquals(-4.0, z.getImaginary(), 1.0e-5);}@*)
		\end{lstlisting}
		\caption{Test cases are sliced into separate code fragments for each assertion. Fragments that depend on production method Complex.negate() are grayed-out.}
		\label{fig2}
	\end{figure}

	\section{Approach}\label{approach}
	\begin{figure}
		\includegraphics[width=\linewidth]{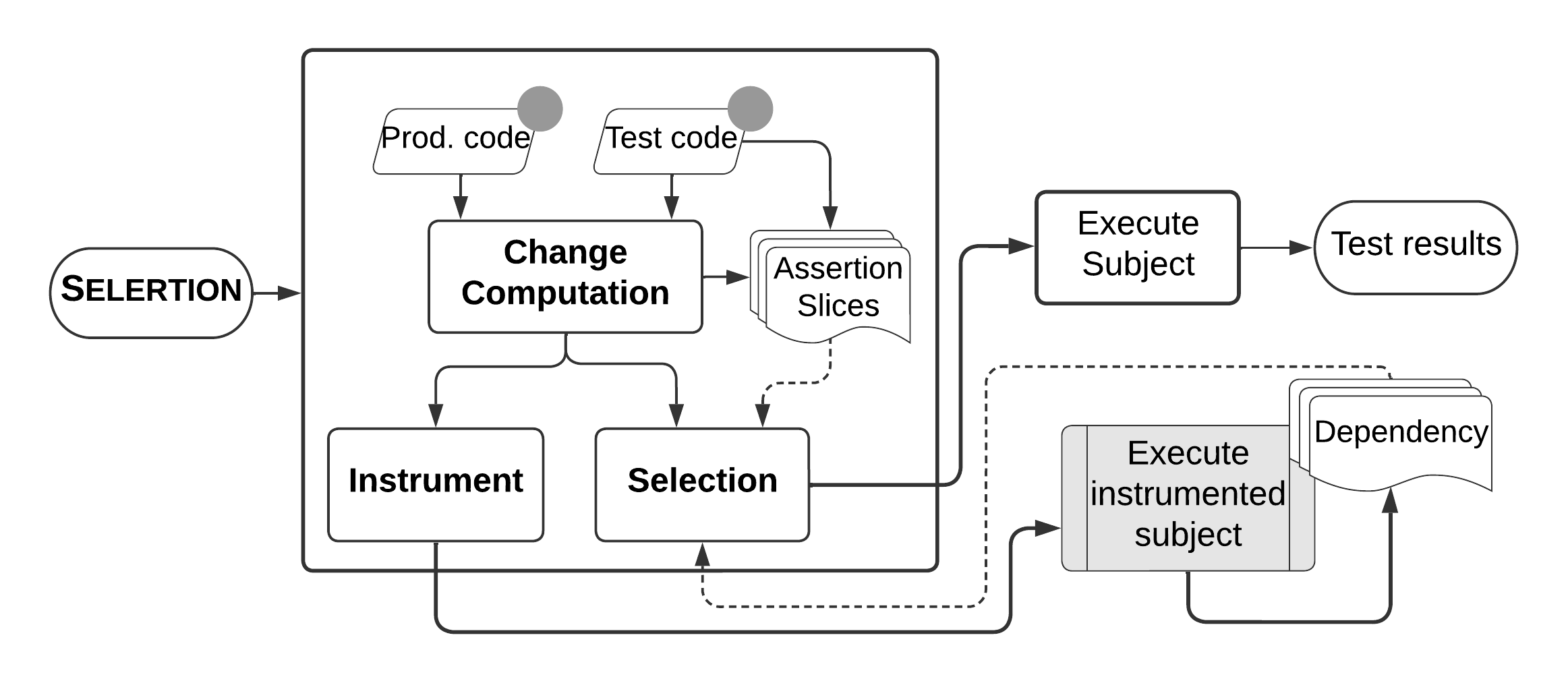}
		\caption{Overview of our approach.}
		\label{overview}
	\end{figure}
	Figure \ref{overview} depicts an overview of our approach, which is composed of four main components. Change computation  detects code changes at the class and method level in the production code. Assertion slicing computes test assertion slices for fine-grained selection by analyzing test code.  Code instrumentation selectively instruments and executes both production and test code to collect dependencies. Finally, test selection selects test entities based on the information collected from the previous components. 
	
	Using the terminology from related work~\cite{gligoric2015practical, zhang2018hybrid}, the analysis (A) phase includes change computation, assertion slicing, and test selection; the collection (C) phase includes the code instrumentation and execution of instrumented subject; the execution (E) phase executes selected test entities. 
	
	\subsection{Change Computation}
	
	\begin{algorithm}
		\small
		{ {\bf Input}: $V_1$, $V_2$: two program revisions}\\
		{ {\bf Output}: $\delta=\delta^c\cup\delta^m$ where $\delta^c$: class-level changes, \newline
			\hspace*{3.5em}$\delta^m$: method-level changes}
		\begin{algorithmic}[1]
			\State{AF,DF,CF = compareFileDiff($V_1$,$V_2$)}\label{ch:1}
			\State{AC,DC,CC = createClassModel(AF, DF, CF)}\label{ch:2}
			\State{$\delta^c$.addAll(AC, DC)}\label{ch:3}
			\ForAll{clz in CC}\label{ch:4}
			\If{$csum(OT)$ or $csum(CH)$ is changed}\label{ch:5}
			\State{$\delta^c$.add(clz)}\label{ch:6}
			\Else\label{ch:7}
			\State{AM, DM, CM = compareMethodDiff(clz)}
			\State{LC = computeLookUpChanges(AM, DM)}\label{ch:9}
			\State{$\delta^m$.addAll(AM,DM,CM, LC)}\label{ch:10}
			\EndIf
			\EndFor\label{ch:12}
			\State{\textbf{return} $\delta=\delta^c\cup\delta^m$}
		\end{algorithmic}
		\caption{Change Computation}
		\label{change}
	\end{algorithm}
	
	Similar to state-of-the-art RTS tools, HyRTS \cite{zhang2018hybrid} and Ekstazi \cite{gligoric2015practical}, we detect code changes by comparing checksums. However, instead of tracing bytecode files, we compute checksums for source code files directly. We perform source code analysis because our fine-grained technique requires computation at statement level for test code to collect dependencies, and statement level analysis for bytecode will be indirect and may increase analysis complexity.
	
	The procedure to compute code changes is summarized in Algorithm \ref{change}. 
	We trace code changes at multiple levels (\textit{i.e.,} method and class level). Given two versions of the code, we first compare the file-level checksums to detect the list of deleted, added, and changed files which are denoted by AF, DF, CF respectively at Line \ref{ch:1}. Then for each file, we create class models 
	to handle circumstances that one single file contains multiple classes (\textit{e.g.,} nested classes), although in most cases, there is one class per file (Line \ref{ch:2}). We also handle \texttt{enum} types. Since enums can be treated as a special type of classes, we use \texttt{class} to represent both types in Algorithm \ref{change}. Intuitively, newly added classes (AC) and deleted classes (DC) are class-level changes, which are collected in the set $\delta^c$ at Line \ref{ch:3}. Changed classes (CC)  may include irrelevant changes that are not impacting the code, such as changes in comments, Javadoc, or indents. Such  changes are excluded at the next step of the Algorithm (Line \ref{ch:4}-\ref{ch:12}). For each changed class model, we separate the class body components into the following categories: class head (CH), methods (M) and other (OT). CH includes the class's name, superclass or interfaces information. Methods (M) include all method declarations, including instance methods and constructors. Other (OT) includes other body declarations such as fields, except methods and classes. We store the smart checksums for each M, CH and OT separately.
	
	The reason to maintain a separate category OT is that, unlike bytecode computation in which field changes are reflected in method-level changes~\cite{zhang2018hybrid}, we need to consider changed fields explicitly in source code analysis. To reduce the analysis overhead, we handle field changes all together and transfer them into a class-level change, because a field may be  used by multiple methods. Therefore, for each changed class, if the OT or CH checksum varies, the corresponding class will be marked as changed (Line \ref{ch:5}-\ref{ch:6}). If there are only method checksum changes without OT or CH change, the added (AM), deleted (DM) and changed methods (CM) will be stored as method-level changes in the set $\delta^m$ (Line \ref{ch:7}-\ref{ch:10}). We also compute look up changes \cite{ren2004chianti, zhang2018hybrid} for added and deleted methods based on the inheritance information at Line \ref{ch:9} for the sake of safety in our analysis. In the initial run, because no previous checksums are available, every class is treated as a new class with class-level changes and the whole test suite is executed once.
	
	\subsection{Test Assertion Slicing}\label{slicing}
	To perform fine-grained selection at the assertion level, we can not merely select the test assertion statement because there could be other test statements associated with the assertion that make it executable. For instance, in Figure \ref{fig1}, the test statement at Line \ref{line:7} is associated with the assertion at Line \ref{line:8}, because the variable \texttt{iPi} is defined at Line \ref{line:7} and used by Line \ref{line:8}. In this work, we call a code fragment that includes the targeted test assertion and the dependent statements as an \textit{assertion slice}.
	
	\begin{definition}[Assertion Slice]
		Let $S_m$ be the list of statements in the test method $m$. An assertion slice encompasses the assertion statement $asrt$, where $asrt \in S_m$ and its dependant statements $S_m[asrt]$, where $S_m[asrt] \subseteq S_m$. 
		
		We use $\mathcal{A}$ to denote the total set of assertion slices for the test suite. $\mathcal{A}_{m}$ denotes a set of slices for the test method $m$. $\mathcal{A}_m[st]$ denotes a set of slices that contain the test statement $st$, where $st \in S_m$. An assertion $asrt$ is also a particular type of test statements.
	\end{definition}
	
	Using Figure \ref{fig2} as an example, the test method ($m$) \texttt{testExp}'s assertion slices $\mathcal{A}_m$ are shown in the first three blocks. For the test statement $st$ at Line \ref{line:3}, $\mathcal{A}_m[st]$ contains one assertion slice shown in the first block. To obtain the assertion slices, we perform static intra-procedural backward slicing \cite{weiser1984program} for each test method based on the \textit{program dependence graph} (PDG) using each assertion and its used variables as slicing criterion \cite{mohapatra2006overview, hammer2004improved, chen2001slicing}. A PDG represents a procedure (i.e., the body of a test method) as a graph where nodes are statements or predicate expressions, and edges are control or data dependence. Generally speaking, test code has simpler structures than production code. Most test cases do not include conditionals such as \texttt{if-else}, \texttt{for-loop}, etc. To reduce the analysis overhead, we opted for only slicing test cases with explicit assertions and without conditional statements. Therefore, we apply data dependency analysis to create the PDG by capturing the variables used and defined for each test statement including the assertions. The global variables and methods annotated with \texttt{@Before} or \texttt{@BeforeClass} in test classes serve a purpose in points-to/alias analysis \cite{hammer2004improved}; should always be included when executing the selected test entities, however, they do not need to be part of the data dependency analysis for individual test cases.
	
	Since we target an object-oriented language (Java), we pay more attention to object variable aliases in our analysis. For instance, when an object variable $\texttt{objA}$ in the assertion statement is selected as the slicing criterion and its reference is passed to a method or constructor of another object variable $\texttt{objB}$, $\texttt{objA}$'s state may be changed by $\texttt{objB}$'s methods~\cite{hammer2004improved, chen2001slicing}. If $\texttt{objA}$  is passed to any field of $\texttt{objB}$,  when $\texttt{objA}$ is changed, $\texttt{objB}$'s state can also change. 
	To make sure the slices do not miss any statements as data dependencies, we make a conservative assumption that method calls will change the states of the caller object and objects passed as arguments, which means a method invocation statement $st$ is associated with the targeted assertion $asrt$ if a variable in $asrt$ is used by $st$ as a caller object or passed to $st$ as an argument. Because assertions are a particular type of statement, if an assertion $asrt_1$ prior to the targeted assertion $asrt_2$ has a data dependency on $asrt_2$, $asrt_1$ will also be included in the slice for $asrt_2$.
	
	\subsection{Code Instrumentation}\label{ci}
	
	After obtaining the assertion slices, we collect test dependencies by capturing the method calls for each test statement. We instrument the production code to log the method signature for each production method invocation, and produce a separate set to keep track of method calls for each test statement inside test cases during test execution. For helper methods that are in test classes but are not test methods (i.e., without \texttt{@Test} annotation), we treat them the same way as production methods.
	
	Performing code instrumentation to collect dependencies at run-time for each new program revision will be inefficient. To reduce the overhead, we instrument each production and test class at initial run $V_0$, and store the instrumented subject as a separate copy (shown as the gray box in Figure \ref{overview}). When a new revision $V_i$ is ready, we only need to re-instrument the changed files computed by Algorithm \ref{change} and synchronize them to the instrumented copy of the subject, and execute the affected tests on the instrumented subject after test selection. Therefore, the dependency collection process is separately executed on the instrumented subject, which does not interfere with the execution of the original subject.
	
	We use $\mathcal{D}$ to denote all the test dependencies. For each test statement $st$, $\mathcal{D}(st)$ stands for a list of method invocations of $st$. Each method call is represented by its fully qualified method signature. Thus, it is easy to retrieve the class-level dependencies when needed.
	
	In practice, some test features might pose safety or efficiency challenges in tracing dependencies for each test statement. For parameterized test classes (annotated with @RunWith(Parameterized.class)), the method calls invoked by each test statement may vary for different inputs. It will increase analysis complexity to trace dependencies for each test statement with consideration of different inputs. We opted for collecting dependencies for each parameterized test class by logging all the method calls for different parameters together into one dependency file. When the dependency file contains any changed methods, all the test cases under different parameters in the parameterized test class are selected for execution. For the test classes that use inheritance, since each subclass may inherit test methods from the super class that are not explicitly available in the subclass, the approach to collect dependencies at test statement level is unsafe because the method-calls invoked by the inherited test methods will be missing for the subclass. Therefore, for the test classes that use inheritance (including both subclasses and superclasses), we trace dependencies at the test class level by logging method calls for the whole test class execution. Moreover, when a test class contains test methods calling other tests, it is hard to isolate each test method. For this case, we collect dependencies at the test class level as well. We identify those test classes that are parameterized, use inheritance or contain test methods calling other tests during static source code analysis for code instrumentation. Then, to trace dependencies at the test class level, instead of inserting code to log method calls for each test statement, we insert code only at the beginning and the end of each test class to keep track of the method calls. 
	
	In addition to these specific test features, some test classes contain test methods that expect an exception to be thrown (e.g., annotated with \texttt{@Test(expected=SomeException.class)}). For those test methods, it is meaningless to trace method calls for each test statement, as the execution will be interrupted at a certain point and statements after that point will not be executed at all. Therefore, we collect dependencies for each test method instead. Moreover, in Section \ref{slicing}, we choose to only slice the tests with explicit assertions and without conditional statements. For the test cases that contain conditional statements such as \texttt{if-else} or \texttt{for-loop} that are not sliced by the assertion slicing module, because they are not selected at fine-grained assertion level, it is sufficient to collect dependencies for each test method instead of each test statement to reduce overhead. 
	
	\subsection{Test Selection}
	\label{sec:testselection}
	
	\begin{algorithm}
		\small
		{ {\bf Input}: $\mathcal{A}$: a set of assertion slices, $\mathcal{D}$: test dependencies, \newline
			\hspace*{2.5em} $\delta=\delta_{P} \cup\delta_{T}$: production and test code changes}\\
		{{\bf Output}: $\Gamma$: selected test entities }
		\begin{algorithmic}[1]
			\State{$\Gamma_{A}, \Gamma_{M}, \Gamma_{C}=\emptyset$}
			\ForAll{{t} in {$\delta_{T}$}}\label{t:1}
			\State{/*update its assertion slices in $\mathcal{A}$*/}\label{t:2}
			\EndFor
			\ForAll{{p} in {$\delta_{P}$}}
			\State{$\mathcal{T}_p$ = retrieveTestEntitiesFrom($\mathcal{D}$, p)}\label{se:3}
			\If{$\mathcal{T}_p \neq \emptyset$ }
			\ForAll{$t$ in $\mathcal{T}_p $}
			\If{$t \in S_m$ }\label{se:6}
			\State{$m = getMethodName(t)$}\label{se:7}
			\If{$\mathcal{A}_m\neq\emptyset$}\label{se:8}
			\State{/*select at test assertion level*/}
			\State{$\mathcal{A}_m[t] = getAssertionSlices(\mathcal{A}_m,t)$}
			\State{$\Gamma_{A}.add(\mathcal{A}_m[t])$}\label{se:11}
			\Else
			\State{/*select at test method level*/}\label{se:13}
			\State{$\Gamma_{M}.add(m)$}\label{se:14}
			\EndIf
			\Else
			\State{/*$t$ is a test method, $\Gamma_{M}.add(t)$ \newline
				\hspace*{6em} or $t$ is a test class, $\Gamma_{C}.add(t)$*/}\label{se:17}
			\EndIf
			\EndFor
			\EndIf
			\EndFor
			\State{$\Gamma = rewriteTests(\Gamma_{A}, \Gamma_{M}, \Gamma_{C}, \delta_{T})$}\label{se:22}
		\end{algorithmic}
		\caption{Test Selection}\label{selection}
	\end{algorithm}
	
	At this point, we have a comprehensive analysis of the test code through assertion slicing and code instrumentation. Based on the levels of dependency data collection and the availability of assertion slices, the test entities will be safely selected at different levels.
	
	\header{Test assertion level} selects related assertion slices inside a test method; test cases that our analysis handles and turns into assertion slices will be selected at this finest granularity level.
	
	\header{Test method level} selects specific test methods in a test class; test cases without any assertion slices or those expecting exceptions to be thrown are selected at test method level.
	
	\header{Test class level} selects all test methods in a test class. Test classes, with special features (as described in Section \ref{ci}), that are parameterized, use inheritance, or contain test cases calling other tests, are selected at the test class level for the sake of safety and efficiency.
	
	Algorithm \ref{selection} summarizes the procedure for our test selection. With the test dependencies $\mathcal{D}$ obtained by running the instrumented subject, the assertion slices $\mathcal{A}$ and the changes $\delta$ computed by Algorithm \ref{change}, the selected test entities $\Gamma$ for the current code revision can be computed by Algorithm \ref{selection}. $\Gamma_{A}$, $\Gamma_{M}$, $\Gamma_{C}$ denote selected assertion slices, test methods and test classes, respectively. We categorize the code changes obtained from Algorithm \ref{change} into two distinct sets: test code changes, denoted as $\delta_{T}$, and production code changes, denoted as $\delta_P$. For each changed test entity $t \in \delta_{T}$ which could be a modified or newly added test method or test class, its assertion slices need to be updated to align with the current code reversion (Line \ref{t:1}-\ref{t:2}). Additionally, $\delta_{T}$ should also be selected for execution (Line \ref{se:22}). For each production code change $p \in \delta_P$, Line \ref{se:3} computes the test entities $\mathcal{T}_p$ that depend on the production code change $p$ based on the dependency files in $\mathcal{D}$. Because there are some special cases that test dependencies are not collected at the test statement level (see Section \ref{ci}), the returned test entities $\mathcal{T}_p$ might contain test methods and test classes. For each affected test entity $t \in \mathcal{T}_p$, if $t$ is a test statement (Line \ref{se:6}), we retrieve the test method $m$ that contains $t$ (Line \ref{se:7}) and check if there are assertion slices available for the test method $m$ (Line \ref{se:8}). If assertion slices exist for $m$, all the assertion slices that contain $t$ are selected (Line \ref{se:8}-\ref{se:11}). If there are no assertion slices for $m$, our selection is performed at the test method level (Line \ref{se:13}-\ref{se:14}) to guarantee a safe selection. If test entity $t$ is not a test statement, $t$ is selected as the test method or test class instead (Line \ref{se:17}). Last but not least, all the selected test entities $\Gamma_{A}$, $\Gamma_{M}$, $\Gamma_{C}$, $\delta_{T}$ are used to rewrite their corresponding test classes to make sure only the selected assertions, methods or classes will be executed (Line \ref{se:22}). After the execution is finished, the test classes are restored to the original status, while the rewritten version is saved as a reference. 
	
	\subsection{Implementation}
	
	We implemented our approach as a tool called \toolname, which is written in Java. For our source code analysis, we build on top of Eclipse Java Development Tools (JDT) \cite{jdt} to analyze and instrument source code. Our current implementation  supports Java programs that use Maven and JUnit 4. In addition to the standard test assertions in JUnit, \toolname supports assertions written in popular frameworks such as AssertJ \cite{assertj} and Google Truth \cite{truth}. At the initial run, \toolname takes the source code of the program, obtains assertion slices, instruments production and test code,  and runs the instrumented program to obtain test dependencies. For any subsequent revision of the program, \toolname takes the new revision of the program as input and computes the changes by comparing checksums with the old revision. It instruments only the changed files and updates the changed assertion slices. It then selects tests at assertion level as the finest granularity where possible by analyzing the dependencies, assertion slices and change information. It then executes the selected tests.
	The code for our implementation is publicly available at \cite{repo}. 
	
	\section{Evaluation}\label{evaluation}
	To evaluate our approach on real-world projects, we target the following research questions:
	\begin{itemize}
		\item \textbf{RQ1 (prevalence):} How prevalent are tests that can be selected at fine-grained test assertion level in practice?
		\item \textbf{RQ2 (precision):} How does our technique compare with existing RTS techniques in terms of precision?
		\item \textbf{RQ3 (efficiency):} How efficient is our approach regarding execution time?
		\item \textbf{RQ4 (effectiveness):} How effective are the selected test cases in detecting faults?
		\item \textbf{RQ5 (performance):} What is the performance of \toolname in terms of analysis overhead and collection time? Is there any performance difference when disabling the assertion-level selection?
	\end{itemize}
	
	\subsection{Existing RTS Tools}
	Our search criteria for existing test selection techniques to compare with included tools that are available, runnable, and target Java/JUnit. We selected two state-of-the-art tools, namely HyRTS \cite{zhang2018hybrid} and Ekstazi \cite{gligoric2015practical}. To the best of our knowledge, these are the exclusive options that currently meet our criteria. Significantly, they outperform older RTS techniques like FaultTracer~\cite{zhang2011localizing}, which exhibit slower performance and offer test method selection, as documented in prior research~\cite{gligoric2015practical}.
	
	HyRTS conducts the analysis of test dependencies at either the \textit{method} or \textit{class} level and selects test entities at the \textit{test class} level. Ekstazi analyzes production code to identify changes at the \textit{class} level. Although the original paper \cite{gligoric2015practical} mentions the provision of both method and class-level selection, the Ekstazi tool \cite{ekstazi} is currently available with default settings that primarily select tests at the \textit{test class} level. Our technique collects test dependencies also at a mix of \textit{method and class} level similar to HyRTS, however, it has the ability to select test entities at multiple levels from the \textit{coarse-grained class to the fine-grained assertion} level based on its analysis of the test suite. Therefore, theoretically speaking, \toolname should be able to select tests more precisely than both HyRTS and Ekstazi. We empirically assess this in our evaluation.
	
	\begin{table*}[!t]
		\caption{Subject systems and their characteristics}
		\label{tab:sub}
        \footnotesize
		\begin{center}
			\begin{tabular}{@{}llrrrrrrrrr@{}} \toprule
				\multirow{2}{*}{Subject} &
				\multirow{2}{*}{Head}&
				\multicolumn{2}{c}{LOC (K)} &
				\multicolumn{2}{c}{Time (s)}&
				\multirow{2}{*}{Test cases} &
				\multicolumn{2}{c}{Fine-grained} &
				\multicolumn{2}{c}{Assertions (\#)} \\ \cmidrule(l){3-4} \cmidrule(l){5-6} \cmidrule(l){8-9}\cmidrule(l){10-11}
				&&Prod. & Test & total & per test & (\#) & (\#) & (\%) & total & per test\\ \midrule
				asterisk-java \cite{asterisk}& 8fe3e4c & 51.4 & 5.0 & 35 & 0.14 & 248 & 187 & 75 & 677 & 2.7\\
				commons-net \cite{net}& 2b0f3383 & 20.0 & 7.6 & 86 & 0.33 & 49 & 29 & 59 & 116 & 2.4\\
				commons-exec \cite{exec}& a374f35 & 1.8 & 1.8 & 99 & 0.95 & 100 & 57 & 57 & 210 & 2.1\\
				tabula-java \cite{tabula} &5f43a93 & 4.6 & 2.3 & 110 & 0.54 & 138 & 94 & 68 & 311 & 2.3\\
				OpenTripPlanner \cite{otp} & 8ee31d5 & 81.1 & 11.4 & 130 & 0.36 & 320 & 225 & 70 & 1,446 & 4.5\\
				commons-math \cite{commonsMath} & c4a093c & 88.4 & 91.3 & 160 & 0.03 & 4,003 & 1,308 & 33 & 11,229 & 2.8\\
				stream-lib \cite{streamlib}&7360389 & 4.8 & 3.8 & 165 & 1.24 & 135 & 38 & 28 & 425 & 3.2\\
				tika-parsers \cite{tika} & fd1926d & 49.5 &21.9 & 181 & 0.16 & 1196 & 221& 18 & 5722 & 4.8\\
				accumulo-core \cite{accumulo}& 5cec6a2 & 193.1 & 26.3 & 338 & 0.32 & 1046 & 433 & 41 & 3325 & 3.2\\
				commons-pool \cite{pool}& 3ca09a7 & 5.5 & 8.9 & 354 & 1.21 & 262 & 28 & 11 & 842 & 3.2\\
				LogicNG\cite{logic}& f94879fa & 23.6 & 12.9 & 546 & 0.67 & 831 & 552 & 66 & 3874 & 4.7\\ \midrule
				Total/Average&-& 523.9 & 193.0 & 2,204 & 0.54 & 8,328 & 3,172 & 48 & 28,177 & 3.3
				\\ \bottomrule
			\end{tabular}
		\end{center}
	\end{table*}

	\subsection{Subject Systems}
	
	We include a total of 11 subject systems. These subjects are selected based on their relevance to prior research on test assertion analysis~\cite{vahabzadeh2018fine, zhang2015assertions} and regression test selection~\cite{zhang2018hybrid} and their popularity on GitHub. To ensure fair comparison among \toolname, HyRTS and Ekstazi, our selection criteria include subjects that use Maven with executable and passing JUnit4 test cases. We focus on subjects with a test runtime of at least 30 seconds. 
	As reported by others~\cite{zhang2018hybrid, gligoric2015practical}, for lower test runtimes it can take more time to analyze the code than to execute the entire test suite. 
	
	The selected subjects and their characteristics are shown in Table \ref{tab:sub} with the ascending order of their testing time. All the subjects are single-module projects, except Tika \cite{tika} and Accumulo \cite{accumulo}. Because some modules of these multi-module projects only have a few test cases, for simplicity, we choose the module with most test cases to analyze (i.e., tika-parsers module for Tika, core module for Accumulo). We examine the test suites for each subject and blacklist test classes that do not meet the selection criteria. Specifically, since our implementation focuses on JUnit4 assertions and identifies test cases with the @Test annotation, we blacklist test classes that mix JUnit3 and JUnit4 APIs for the OpenTripPlanner subject. For the tika-parsers subject, we choose to skip test classes that mix Mockito \cite{Mockito} and JUnit4 APIs. Additionally, 9 out of the 11 subject systems, except for tabula-java and stream-lib, have recently migrated from JUnit4 to JUnit5 \cite{junit5}. To maintain consistency, we start from a revision that uses JUnit4 as the Head to collect 100 revisions before it, and remove revisions that fail to build or have no code changes. For the three subjects commons-exec, commons-net and commons-pool, which lack a sufficient number of revisions with substantial source code changes, we introduce changes artificially. Because \toolname detects production code changes at a coarse-grained class or method level, our focus is on the location of the changes rather than the specifics of the changes themselves. By leveraging generated mutants, we can introduce code changes at various locations within the production code. We believe that these mutants can effectively mimic real code changes in this context. To achieve this, we use a source code mutation generator called Major \cite{major} to create random mutants in various production methods, thus introducing more prevalent source code changes. The selected subjects span different application domains and exhibit diverse code and test suite sizes. Table \ref{tab:sub} presents the short hash under the Head column, and lines of code for both production and test code for the Head revision. We measure lines of code using Cloc \cite{cloc}. The column Time details the average time required to execute all the tests using the build command (i.e., \texttt{mvn test}) for the revisions we analyzed, along with the average execution time for each test case. This table also presents information pertaining to test cases and test assertions that are further elaborated in the following subsection.
	
	\subsection{Procedure and Results}
	All our experiments are conducted on a macOS server with 2.4 GHz Quad-Core Intel Xeon processor and 64 GB of memory running OpenJDK 64-Bit version 1.8.0\_212 and 11. 
	
	\subsubsection{Prevalence (RQ1).} 
	
	Given that we exclude parameterized tests, inherited tests, and tests dependent on other tests from our assertion-level selection (such tests are selected at method or class level), with consideration for the safety of our analysis, it is important to know to what extent the test suite for each subject can be fine-grained selected at the assertion-level. To measure the prevalence of fine-grained selection, we implemented a tool using JDT \cite{jdt}. This tool statically counts the total number of test cases that can be fine-grained selected for every revision we executed. We present the average values for each subject in Table \ref{tab:sub}. Additionally, because the smallest unit of interest for our study is the test assertion, we count the total number of assertions and calculate the average number per test. These numbers are included under the Assertions column in the table, both in total and per test. It's worth noting that we opt to count these numbers statically instead of relying on information obtained from executed tests (e.g., as provided by maven-surefire-plugin \cite{surefire}) to avoid repetitive counting, as a single test case may be executed multiple times due to inheritance or different parameters.
	
	The column Fine-grained in Table \ref{tab:sub} shows the number and percentage of test cases that can be fine-grained selected by \toolname in the subject systems. The results show on average 48\% out of the total 8,328 test cases can be selected at the assertion level. The percentages range from 11\% (commons-pool) to 75\% (asterisk-java). Moreover, if each test case contains less than two assertions, the benefits of fine-grained selection become limited because the assertion-level selection will be equivalent to the method-level selection (i.e., the whole test method is selected). Therefore, we count explicit assertion statements (e.g., \texttt{assertTrue} ) inside each test method body. Column Assertions in Table \ref{tab:sub} shows the number of assertions for each subject, which totals 28,177 for all subjects. The average number of assertions per test case ranges from 2.1 for commons-exec to 4.8 for tika-parsers, and the overall average of all the subjects is 3.3 assertions per test case. This indicates that (1) most test cases contain multiple assertions in practice, (2) tests that are analyzable by our approach at the fine-grain assertion-level are prevalent, and (3) our fine-grained selection approach is feasible to be applied to real-world test suites.

	\subsubsection{Precision (RQ2).} 
	
	To assess the efficacy of our approach in terms of precision, we use the widely used RTS metric \textbf{Selected Test Ratio} \cite{harrold2001regression, gligoric2015practical, legunsen2016extensive, orso2004scaling, ren2004chianti, zhang2011localizing, zhang2018hybrid} to compare with the other two RTS tools. \textbf{Selected Test Ratio} is defined as the ratio of the number of selected test cases to the number of total test cases. As the fine-grained selection is at the test assertion level in this study, we define a new metric \textbf{Selected Assertion Ratio} as the ratio of the number of selected assertions to the number of total assertions. Lower selected test/assertion ratio indicates better selection precision, as it signifies the selection of fewer tests.
	
\begin{figure}
		\centering
		\includegraphics[width=0.9\linewidth]{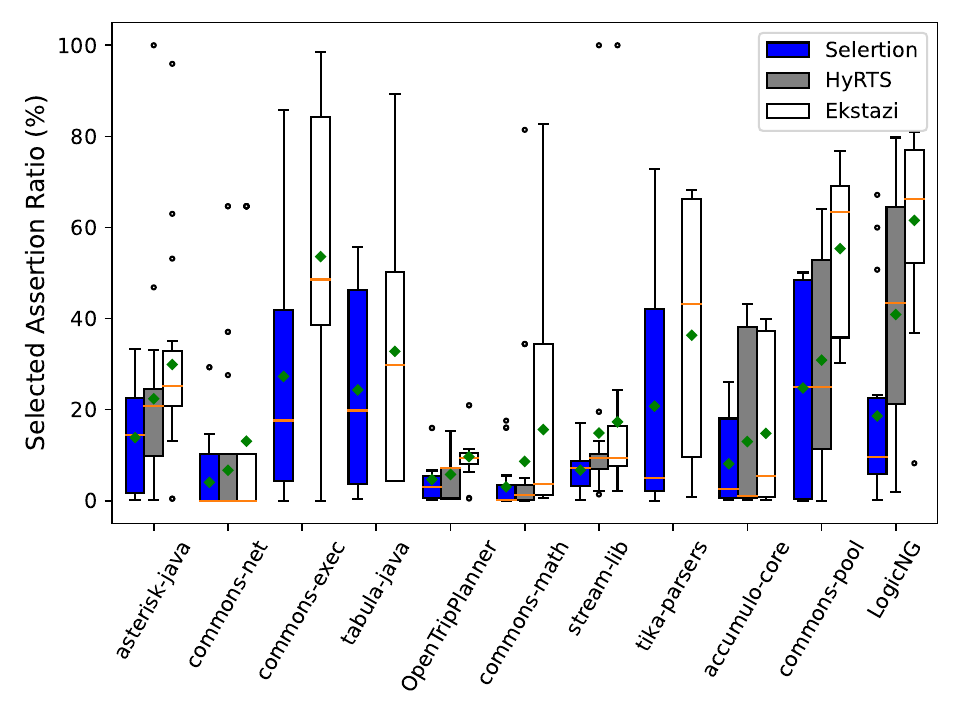}
		\caption{Distributions of selected assertion ratio. Lower values are better as fewer tests are selected.}
		\label{selectedAssertions}
	\end{figure}

	Our precision evaluation results are presented in Table \ref{tab:ratio} and Figure \ref{selectedAssertions}. In Table \ref{tab:ratio}, the macro-columns Selected Assertion Ratio and Selected Test Ratio present the mean values of all the revisions we studied for \toolname, HyRTS and Ekstazi, respectively, except for commons-exec, tabula-java and tika-parsers. In the case of these subjects, HyRTS failed to run due to StackOverflow errors. Therefore, we mark the failed subjects and the corresponding average values for HyRTS with an asterisk to indicate that these values are calculated by excluding the subjects that failed, resulting in an average calculation different from that of Ekstazi. For a fair comparison, we calculate two average values for \toolname: one is the average across all subjects for comparison with Ekstazi, and the other, marked with an asterisk, is the average value excluding the HyRTS-failed subjects for comparison with HyRTS. As our results show, \toolname has lower selected test ratio values, which means it selects fewer test cases and assertions than the other two tools. Regarding selected assertion ratio, on average, \toolname selects 15.10\% out of all the test assertions and all subjects, which is 52.1\% less than Ekstazi. When compared with HyRTS (excluding the marked subjects), the average of \toolname is 10.63\% which is 40.8\% less than HyRTS's average of 17.96\%. For selected test ratio, overall our tool selects 15.80\% of all the test cases (10.85\% when excluding the marked subjects). This is 43.4\% less than HyRTS and 53.0\% less than Ekstazi. For each subject, the precision increase in terms of selected assertion ratio is from 10.8\% (OpenTripPlanner) to 65.1\% (commons-math) in comparison with HyRTS, and from 16.1\% (tabula-java) to 80.8\% (commons-math) compared with Ekstazi. Meanwhile, our tool decreases the selected test ratio from 22.9\% (commons-pool) to 94.4\% (commons-net) in comparison to HyRTS, and from 18.2\% (tabula-java) to 94.9\% (commons-net) compared with Ekstazi. It is worth mentioning that the selected ratios are related to the size of code changes in each sample revision we collect to analyze. Larger code changes always lead to higher selected ratios. Moreover, to provide more insights about selected assertion ratios for each subject besides the mean values, Figure \ref{selectedAssertions} includes boxplots depicting the distributions of selected assertion ratios for each revision. The mean values are marked as diamonds, and median values are marked as lines. The results indicate that our approach has the ability to select tests more precisely than other coarser-grained tools.
	
	\begin{table}[!t]
		\caption{Results: selected test and assertion ratios}
		\label{tab:ratio}
        \footnotesize
		\begin{center}
			\setlength\tabcolsep{1pt} 
			\begin{tabular}{@{}l@{\hspace{-6pt}}cccccr@{}} \toprule
				\multirow{2}{*}{Subject} &\multicolumn{3}{c}{Selected Assertion Ratio (\%)}
				&\multicolumn{3}{c}{Selected Test Ratio (\%)}
				\\ \cmidrule(l){2-4}\cmidrule(l){5-7}
				& \toolname & HyRTS & Ekstazi & \toolname & HyRTS &  Ekstazi \\ \midrule
				asterisk-java &	13.9&	22.4&	29.9&	12.9&	20.6&	26.7\\
				commons-net & 4.0 & 6.7 & 13.1 & 0.3 & 5.4 & 5.9 \\
				commons-exec* & 27.3 & - & 53.6 & 28.4 & - & 53.9 \\
				tabula-java* & 32.8& - & 39.1 &	39.6&	-&	48.4\\
				OpenTripPlanner & 5.8& 6.5&	10.4&	5.9&	9.8&	16.8\\
				commons-math &	3.0&	8.6&	15.6&	7.9&	11.9&	19.6\\
				stream-lib & 6.9&	14.7&	17.0&	11.0&	19.3&	22.0\\
				tika-parsers* &  21.0 & - &36.2 & 19.0 & - & 32.9\\
				accumulo-core & 8.1& 13.0&	14.9&	3.7&	10.0&	11.4\\
				commons-pool & 24.7 & 30.9 &	55.4 & 29.3 & 38.0 & 71.5 \\ 
				LogicNG & 18.6 & 40.9 & 61.5 & 15.8 & 38.3 & 60.3 \\\midrule
				Ave. & 15.10 / 10.63* & 17.96* & 31.52 & 15.80 / 10.85* & 19.16* & 33.58
				\\ \bottomrule
			\end{tabular}
		\end{center}
	\end{table}
	
	\subsubsection{Efficiency (RQ3).} 
 
 Because our approach is implemented in an offline fashion, where the data collection phase is executed in a separate process and the instrumented source code is stored in a separate copy of the original project, we measure the \textit{offline End-to-End time} (AE time) for \toolname, HyRTS, and Ekstazi, respectively, to ensure a fair comparison and assess the efficiency in reducing testing time. The offline End-to-End time for \toolname includes the analysis (A) time to compute code changes, select affected test entities and update assertion slices, as well as the execution (E) time for the selected test entities, then normalize the end-to-end time into a ratio with respect to the overall time of executing the whole test suite (RetestAll). We perform the measurement for each revision and report the average. We also measure the offline end-to-end time and its ratio for HyRTS and Ekstazi respectively. The results are shown in Table \ref{tab:time}. Column  RetestAll shows the time required to run the whole test suite in average for each subject as a reference. The other three macro-columns present the offline end-to-end time and its ratio to RetestAll for \toolname, HyRTS, and Ekstazi separately. Besides the average values in Table \ref{tab:time}, to further understand the results, we perform Paired T-Test~\cite{marascuilo1988statistical} with the significance level as 5\% ($\alpha$ = 0.05) to determine if there is a significant difference between each two groups (i.e., \toolname vs.\ HyRTS,  \toolname vs.\ Ekstazi and HyRTS vs.\ Ekstazi) for all the revisions of every subject. The reason to choose Paired T-Test is that as a parametric test, it has greater statistical power than non-parametric tests like Wilcoxon Test \cite{marascuilo1988statistical}, and can provide trustworthy results even for non-normal distributions. If the end-to-end time has no statistical difference in comparison with others, it means it has competitive or equivalent capability for reducing test time. After examining the results of offline end-to-end time, in Table \ref{tab:time}, we mark the subject as light gray if \toolname performs competitively or better than one of the other tools; otherwise, \toolname outperforms the other two tools for that subject. We also highlight the shortest average running time in yellow in each row across the three tools. 
	
	Across all 11 subjects, \toolname consistently reduces testing time effectively, with ratios to RetestAll ranging from 14.9\% (commons-net) to 70.3\% (OpenTripPlanner). This reduction translates to an overall decrease in testing time of 29.7\% to 85.1\%. For asterisk-java with 35 seconds runtime, our tool is competitive with HyRTS, but slower than Ekstazi. For subject commons-net, although HyRTS has the shortest running time, \toolname is only 1 second slower than HyRTS and is equivalent to Ekstazi, which has no statistical difference in comparison with both tools. For subject commons-math, \toolname outperforms Ekstazi but is 5.3 seconds slower than HyRTS.

	\header{Longer Test Runtime.}
	For the subjects with runtimes longer than 86 seconds, \toolname performs significantly better than Ekstazi in terms of testing time. HyRTS fails to run on the subjects commons-exec, tabula-java and tika-parsers, thus we treat \toolname as superior on these three subjects. As an exception, commons-math has long test time with 160 seconds, the reason that HyRTS works better than \toolname is the execution time per test for commons-math is very fast with only 0.03s (shown in Table \ref{tab:sub}). Therefore, although test assertions are reduced significantly, the reduced execution time is marginal. Although HyRTS has lower average testing time for commons-math (row 8), the result of Paired T-Test indicates that there is no significant difference between \toolname and HyRTS. \toolname performs better than HyRTS for the remaining subjects. Generally speaking, we can conclude that Ekstazi works best for subjects with short test runtime such as asterisk-java with 35s running time, \toolname outperforms or is competitive to both HyRTS and Ekstazi with long-running subjects i.e., test runtime longer than 86 seconds in our experiments. The results show that our fine-grained RTS technique is more effective in handling time-intensive tests. 
	
	Considering all test suites in our subjects, short and long, the total end-to-end time on average for \toolname is 74.33 seconds while the average overall test execution time (RetestAll) is 200.36 seconds. To the ratio of RetestAll, our technique reduces 63\% of the overall test execution time. Compared with Ekstazi which has an average end-to-end time of 119.74 seconds, \toolname is overall 38\% faster than Ekstazi. Compared with HyRTS when we ignore commons-exec, tabula-java and tika-parsers, the average end-to-end time for HyRTS is 83.51 seconds and 77.51 seconds for \toolname which is 7\% faster than HyRTS. The average gain compared with HyRTS could be  potentially larger if HyRTS were executable on commons-exec, tabula-java and tika-parsers. Our results show that \toolname is more precise and more efficient, especially for test suites with longer execution times, which benefit most from test selection in practice. 
	
	\begin{figure*}
		\begin{minipage}[c]{0.55\textwidth}
			\captionof{table}{Results: testing time and comparison with other RTS tools}
			\label{tab:time}
			\footnotesize
			\setlength\tabcolsep{2pt} 
			\begin{tabular}{@{}lcccccccc@{}} \toprule
				\multirow{2}{*}{Subject} & \multirow{2}{*}{RetestAll}
				&\multicolumn{2}{c}{\toolname}
				&\multicolumn{2}{c}{HyRTS}&	\multicolumn{2}{c}{Ekstazi}
				\\ \cmidrule(l){3-4}\cmidrule(l){5-6}\cmidrule(l){7-8}
				& s& s & \% & s & \% & s & \% \\ \midrule
				\rowcolor{LightGray}
				asterisk-java & 35 & 19.4 & 58.7& 18.1 & 53.6 & \cellcolor{bananayellow}11.7 & 35.0\\
				\rowcolor{LightGray}
				commons-net & 86 & 13.0 & 14.9 & \cellcolor{bananayellow}11.9 & 13.7 & 12.9 & 14.9 \\
				commons-exec* & 99 & \cellcolor{bananayellow}48.6 & 49.1 & - & - & 70.2 & 70.9 \\
				tabula-java* & 110 & \cellcolor{bananayellow}76.1& 69.3 & - &- & 81.0 & 73.9\\	
				OpenTripPlanner & 130 & \cellcolor{bananayellow}89.8& 70.3 & 93.5 &73.1 & 119.9 & 94.0\\
				\rowcolor{LightGray}
				commons-math & 160& 49.0 & 30.5 & \cellcolor{bananayellow}43.7 & 27.2 & 63.2 & 39.4\\			
				stream-lib & 165 & \cellcolor{bananayellow}42.4& 25.7& 52.1 & 31.4 & 53.9 & 32.5\\		
				tika-parsers* & 181 & \cellcolor{bananayellow}72.8& 40.0 & - & - & 80.3 & 44.3\\	
				accumulo-core & 338 & \cellcolor{bananayellow}66.7& 19.7& 80.1&23.6 &104.3 & 30.8\\ 
				commons-pool & 354 & \cellcolor{bananayellow}160.7 & 45.4 & 176.7 & 49.9 & 308.0 & 87.0\\ 
				LogicNG & 546 & \cellcolor{bananayellow}179.1 & 32.8 & 192.0 & 34.7 & 411.7 & 74.9\\
				\midrule
				Ave.&200.36&74.33 / 77.51* & 41.49 / 37.25* &83.51*& 38.40* &119.74& 54.33
				\\ \bottomrule
			\end{tabular}
		\end{minipage}
		\hfill
		\begin{minipage}[c]{0.4\textwidth}
			\captionof{table}{Killed mutants by the original test suite and RTS tools}
			\label{tab:FDA}
			\footnotesize
			\setlength\tabcolsep{2pt} 
			\begin{tabular}{l|c|ccc} \toprule
				Subject &(Original)&	\toolname&HyRTS&Ekstazi\\
				\midrule
				asterisk-java  &2&2&2&2	\\
				commons-net  &3&3&3&3 \\
				commons-exec* &2&2&-&2 \\
				tabula-java*&3&3&-&3\\
				OpenTripPlanner&1&1&1&1\\
				commons-math&5&5&5&5\\
				stream-lib &5&5&5&5\\
				tika-parsers* &3&3&-&3\\
				accumulo-core &5&5&5&5\\
				commons-pool &2&2&2&2\\
				LogicNG &5&5&5&5\\
				\bottomrule
			\end{tabular}
		\end{minipage}
	\end{figure*}
	\subsubsection{Effectiveness (RQ4).} \label{RQ4} 
 
 A safe RTS tool should ensure that the number of newly failed tests when run with the tool matches the number when running the original test suite \cite{zhu2019framework}. To asset the effectiveness of \toolname and compare it with state-of-the-art approaches, we employ Major \cite{major} to randomly generate ten mutants across the production code for each subject and then count the number of eliminated mutants for the original test suite, \toolname, HyRTS and Ekstazi. The results are reported in Table \ref{tab:FDA}. For all subjects, the total number of mutants killed, as well as each individual mutant eliminated by the RTS tools \toolname, HyRTS and Ekstzai, align with those killed by the original test suite. These empirical findings affirm that our fine-grained selection approach maintains the same test effectiveness as observed in the original test suite, demonstrating that it is safe. Additionally, we discuss other efforts taken to ensure the safety of \toolname in Section \ref{discussion}.

	\subsubsection{Performance (RQ5).} 
	
	\begin{table*}[!t]
		\caption{Performance results for \toolname and comparison to a method-level variant}
		\label{tab:compare}
        \footnotesize
		\begin{center}
			\setlength\tabcolsep{4pt} 
			\begin{tabular}{@{}lccccccccllllll@{}} \toprule
				\multirow{3}{*}{Subject}&\multicolumn{3}{c}{End-to-end Time (AE)}
				&\multicolumn{3}{c}{Analysis Time (A)}&\multicolumn{2}{c}{Collection Time (C)}  &\multicolumn{2}{c}{Initial Overhead} &\multicolumn{4}{c}{Selected Assertions}
				\\ \cmidrule(l){2-4}\cmidrule(l){5-7}\cmidrule(l){8-9}\cmidrule(l){10-11}\cmidrule(l){12-14}
				& Asrt & Mthd & diff & Asrt & Mthd & diff& Asrt & Mthd & Asrt & Mthd & Asrt & Mthd & diff\\ 
				& s (\%) & s & s & s (\%)&s & s & s & s & s & s & \% & \% & \% \\\midrule
				asterisk-java &19.36	(58.74)&17.25 &2.11 &3.84 (11.72)&3.71 &0.14 &17.51	&17.40	&34.26	&33.30 &13.87	&14.24 &-0.37 \\
				commons-net & 12.96 (14.89) & 12.86 & 0.10 & 1.17 (1.34) & 1.16 & 0.01 &12.31	&12.22	&89.86	&95.21 &4.03	&4.95 &-0.92 \\
				commons-exec & 48.64 (49.13) & 48.76 & -0.12 & 2.56 (2.59) & 2.05 & 0.51 &55.89 &57.14	&143.14	& 101.77&27.29	&28.91 &-1.62 \\
				tabula-java &76.07 (69.32)&75.97 &0.10 &6.24 (5.69)&6.15 &0.09 &76.71&77.14 &122.02 &122.05&32.80 &32.96 &-0.15 \\
				OpenTripPlanner &89.82 (70.33)&89.92 &-0.10 &13.60 (10.68)&13.32 &0.28 &142.76 &143.49 &640.88 &659.89&5.76 &6.22 &-0.46 \\
				commons-math &49.00	(30.55)&48.11 &0.89 &6.55 (4.08)&5.93 &0.62 &108.23&107.38	&617.13	&689.03&3.04 &3.05 &-0.01\\
				stream-lib &42.43	(25.70)&41.64 &0.79 &2.25 (1.36)&2.18 &0.07 &108.38 &107.80&367.89	&363.19&6.90	&6.96 &-0.06\\
				tika-parsers &72.80 (40.04)&71.22 &1.57&18.93 (10.48)&18.14 &0.80 &108.55	&99.88	&887.99	&926.05&21.00&21.02&-0.02\\
				accumulo-core &66.70 (19.71)&78.18 &-11.48 &5.51 (1.63)&4.91 &0.60 &59.49	&96.87	&585.35	&580.23&8.13 &8.22 &-0.09\\
				commons-pool & 160.73 (45.40) & 168.80 & -8.07 & 2.36 (0.67) & 2.50 & -0.14 &173.58	&187.88	&403.94	&401.76 &24.74	&26.74 &-2.00\\
				LogicNG &179.06 (32.78) & 175.77 & 3.29 & 12.08 (2.20) & 9.05 & 3.03 & 744.54 & 767.02 & 6216.11 & 6207.72 &18.60 & 18.86 & -0.26\\
				\midrule
				Ave.&74.32&75.32 &-0.99 &6.83 & 6.28 &0.55 & 146.18 & 152.20 & 918.96	& 925.47 & 15.11 & 15.65 &-0.54\\ \bottomrule
				
			\end{tabular}
		\end{center}
	\end{table*}
	
	To evaluate the performance of \toolname, we measure the offline end-to-end time (AE time), analysis time (A time), collection time (C time) for dependency collection which runs in a separate process, as well as the initial overhead, respectively. In addition, we disable the finest-grained selection component of \toolname  to assess the impact of  selecting test entities at the assertion level. After disabling the component, the granularity of selection becomes test method level. To have a safe analysis, selection at the test method level still requires instrumenting the test code at statement level to inspect whether there exists any test method that depends on other tests. As before, the test classes that contain test methods calling other tests are selected at test class level along with those using inheritance or parameterized tests.
	
	For the AE time, we already presented the results for \toolname in Table \ref{tab:time}. To compare with the method level selection, we present the results in seconds and percentages in Table \ref{tab:compare} under End-to-end Time (AE). We report  the difference for AE time between \toolname and the method-level variant. Our analysis reveals that, when comparing assertion-level selection to method-level selection, there is some variation in end-to-end time across different subjects. For subjects like commons-exec, OpenTripPlanner, accumulo-core, and commons-pool, assertion-level selection consistently requires less end-to-end time. However, for the remaining subjects, assertion-level selection tends to demand slightly more time, with differences ranging from 0.1 second (as seen in commons-net) to 3.29 seconds (observed in LogicNG) on average. Notably, the overall average for time difference is -0.99 seconds, indicating that assertion-level selection is, on average, faster. To gain further insights from our results, we conduct a Paired T-Test ($\alpha = 0.05$) to compare the performance of \toolname with its coarser-grained variant. The analysis reveals that, for the first group of subjects, including tabula-java, accumulo-core, and commons-pool, \toolname is significantly faster than the method-level selection, with a p-value of 0.041, which is lower than $\alpha$. In contrast, for the second group, which includes the remaining eight subjects, there is no statistically significant difference in the offline end-to-end running time between \toolname and its coarser-grained variant. Therefore, we can confidently conclude that \toolname is either significantly faster or equivalent to its method-level variant.
	
	The analysis time  (A) is positively correlated to the size of code changes, because large code changes require more time for change computation. For example, because the revisions of tika-parsers have relatively large code changes, as shown in Figure \ref{selectedAssertions} where the boxes for the subject are higher than the others, tika-parsers has the longest analysis times of 18.93 seconds. However, for the long-running subjects, the analysis time only takes a small portion of the overall time. Tika-parsers has 18.93 seconds of analysis time on average which is only 10.48\% of the RetestAll. In contrast, asterisk-java has 3.84 seconds analysis time which is much shorter than 18.93 seconds, but the analysis time is 11.72\% of the RetestAll. That might also explain why \toolname works better on long-running subjects. To compare with its method-level variant, from the results in the columns under Analysis Time (A) in Table \ref{tab:compare}, the analysis time for \toolname is more than method-level selection for each subject. Overall, it takes 0.55 second more on average than method-level selection for the analysis phase. Besides the mean values reported in Table \ref{tab:compare}, we also calculate the Paired T-Test. The p-value is 0.015 which is much less than $\alpha$ (i.e., $p < \alpha$, where $\alpha = 0.05$); thus, we conclude that fine-grained assertion selection requires more analysis time than method-level selection, as expected. 
	
	In our implementation, the collection phase operates as a separate process, allowing developers to obtain RTS results without needing to wait for the data collection to complete. However, it is still important to measure collection time to evaluate performance. The results for collection time are presented under Collection Time (C) in Table \ref{tab:compare}. The collection time varies among subjects. For some subjects, including asterisk-java, commons-net, accumulo-core, commons-exec, tabula-java, and commons-pool, the collection time is either shorter than or comparable to the AE time. However, for a subject like LogicNG, the collection time is significantly longer. Upon manual inspection of the LogicNG codebase, it was found that the test cases include heavy use of for loops. We plan to improve the instrumentation for test cases with extensive loop executions in the future. Nonetheless, our implementation decouples the collection phase from the test selection approach, allowing developers to control when and whether to trigger the collection process. This process runs in the background on a separate copy of the codebase, ensuring it does not interfere with the developer's work on the original codebase. Considering all the 11 subjects as a whole, on average, the collection time is 146.18 seconds for \toolname, while 152.20 seconds for method-level selection. However, based on the Paired T-Test, the collection time for method level selection has no statistical difference with \toolname.
	
	The initial overhead includes code instrumentation, backward assertion slicing and dependency collection at the initial run. In Table \ref{tab:compare}, we present the results under Initial Overhead. For assertion-based selection, the average time consumption is 918.96 seconds (15.3 min). Similarly, the result of Paired T-Test indicates there is no statistical difference for method level selection in terms of initial overhead. Because the initial overhead is just a one-time requirement, it is acceptable compared to the execution time saved by our tool over time. 
	
	Besides the time components, we also compare the selected assertion ratio between \toolname and the method-level variant. The results are shown in the fifth macro-column in Table \ref{tab:compare}. Overall, fine-grained test selection will select fewer assertions than method-level selection. On average, \toolname chooses 0.54\% fewer assertions per revision for each subject, which translates to 14 fewer assertions in absolute numbers. The p-value of Paired T-Test is 0.006 for selected assertion ratio indicates that the fine-grained selection is significantly better than method-level selection in terms of selected assertion ratio.
	
	After comparing with the variant by disabling fine-grained assertion selection, although \toolname requires more analysis time, its end-to-end time, collection time and initial overhead are either less or equivalent to the coarse-grained variant. This indicates that fine-grained selection generally does not consume more time or result in a performance loss. Moreover, fine-grained assertion-level selection is either significantly faster or equivalent to its method-level variant and exhibits the ability to select fewer assertions with higher precision across all subjects. Therefore, we can conclude that fine-grained assertion-level selection surpasses method-level selection in terms of overall performance and precision.
	
	\section{Discussion}\label{discussion}
	In this section, we discuss our efforts to guarantee safety for the fine-grained selection, reflect on our findings, tool design decisions and limitations, as well as threads to validity of our study.
	
	\header{Fine-Grained Selection Safety.} In addition to the analysis in Section \ref{RQ4}, where we leveraged mutation testing to evaluate the test effectiveness across all subjects, we manually introduced real bugs from the Defects4J v2.0.1 \cite{just2014defects4j} database. Upon inspection, we found that the test suites for certain projects in Defects4J, such as Chart and JacksonCore, are still using JUnit3, while our tool is designed to support JUnit4. As a result, we chose to assess our tool with real bugs in commons-math, the project common to both our subject systems and Defects4J. \toolname successfully selected the same triggered tests as those documented in Defects4J for all the bugs. Regarding our approach to production code analysis, we follow the design principles of prior works, such as HyRTS \cite{zhang2018hybrid} and Ekstazi~\cite{gligoric2015practical}; both argue they are safe in their test selection. Our analysis granularity for production code is a combination of method and class level, because we transform changes outside method bodies to coarse-grained class-level changes. It has been proven by Zhang~\cite{zhang2018hybrid} that this kind of change transformation cannot introduce new safety issues. Additionally, for inheritance in production code, since the inheritance relationship is captured during code instrumentation, if changes in parent class A impact child class B, all test entities related to both classes A and B will be selected.
	
	Thus, if there are any potential safety issues in our technique, they may derive from the fine-grained instrumentation and selection for test code. Our assertion slicing component (see Section \ref{slicing}) determines whether a test can be sliced into assertion slices. We have performed many experiments to make sure the assertion slices are executable without breaking the original test methods. To guarantee a safe selection, we perform specific analysis (see Section \ref{ci}) to discreetly identify tests that are unfeasible for finer-grained selection including both test method and assertion level and take a conservative selection approach (see Section \ref{sec:testselection}). 
	For tests containing statements out of our analysis scope we also take a conservative approach to be safe. For instance, test statements that call third-party libraries such as \texttt{Thread.sleep}, are not safe to be selected at assertion level, because the statement \texttt{Thread.sleep} will not be included in any assertion slices. We downgrade such tests to method-level selection to be safe.
	
	\header{Source code vs.\ Bytecode.} Different from the other tools that identify changes using Java bytecode, we focus on changes at source code level. We would like to mention that there are some scenarios that may lead to completely different results. For example, if the program revision is to upgrade the Java version without any real source code changes, our tool will not select any tests, however, because the bytecode is changed completely, the other tools will select the whole test suite. In cases where configuration changes occur without any real source code changes, developers might choose to execute the entire test suite without relying on an RTS tool. Moreover, some modifications of source code do not impact the bytecode, such as changing \texttt{a+=1} to \texttt{a=a+1}; here \toolname selects  tests related to the source code changes, while the other Bytecode analysis based tools will not select any tests. We believe that our fine-grained, source code analysis-based tool is more suitable for developers to verify their source code changes in daily tasks, whereas bytecode-based tools are better suited for use in a continuous integration environment. 
	
	\header{Applications.} Besides regression test selection, our technique can be useful for developers in understanding  and maintaining test code. Developers often have to take over other developers' tasks. Unit tests are a great resource to understand the existing code. With the help of our tool, developers can make some small changes in a method without changing its functionality to obtain detailed test entities, especially test assertions related to the method, and use the information to understand the method and its relation with the test. The assertion slicing module can also be used to detect test smells such as redundant statements in test cases. Currently, our implementation only targets unit tests to compare with the prior works. However, unit tests usually run fast; for example, the average execution time per assertion is only around 0.08 seconds for the subjects in our experiments. Our fine-grained selection technique can be more powerful on tests with longer running time, such as tests in the web and mobile app domains.
	
	\header{Limitations.} Our implementation currently supports assertions from JUnit4 and assertion frameworks built on top of JUnit4 such as Google Truth \cite{truth} and AssertJ \cite{assertj}. However, the tool provides flexibility to manually add a list of assertions (customized or from other frameworks) to be included in the analysis. We do not support fine-grained selection for multithreaded tests, because the strategy for dependency collection is designed to collect method dependencies for each test statement in a sequential fashion which might not work for multithreaded tests.
	
	\header{Threats to Validity.} The main threat to internal validity is that our implementation may have bugs that may impact the results. To improve the confidence for our implementation, we performed numerious tests to make sure the selection results do not break original test cases, and also manually inspected the experiment results for several subjects. Using a limited number of subject systems and mutants to simulate code changes for only three out of the 11 subjects in our evaluation could pose an external validity of our results to be generalized. We tried to minimize the threat to select subjects from different domains, with various lines of code and wide range of test execution time. Moreover, the subjects selected are well-maintained and widely used in recent RTS \cite{zhang2018hybrid} and fine-grained analysis \cite{vahabzadeh2018fine} work. We opted to use mutants to simulate code changes for the three subjects that lack real code changes, because our analysis of production code is coarse-grained, focusing on the location of changes rather than the specifics of the changes themselves. To enhance the representativeness of these simulated code changes, we introduced mutants randomly into various locations within the source code to closely mirror real-world situations. 
	
	\section{Related Work}\label{related}
	
	Regression test selection (RTS) has been a widely studied topic, with various techniques ranging from basic-block-level to module-level granularity reviewed in several surveys \cite{engstrom2008empirical, engstrom2010systematic, biswas2011regression, yoo2012regression, kazmi2017effective}. The basic-block-level RTS, introduced by Rothermel and Harrold for C programs and later extended to Java \cite{rothermel1997safe, harrold2001regression}, evolved into coarser-grained methods to improve efficiency. Notable among these are \textsc{FaultTracer} and \textsc{Chianti} at method level \cite{zhang2011localizing, ren2004chianti}, and \textsc{Ekstazi} at file level \cite{gligoric2015practical}, with \textsc{Ekstazi} demonstrating superior performance in testing time and scalability. Further research explored coarser granularities, like file-level and module-level RTS \cite{vasic2017file, celik2017regression}. The trade-off between precision and efficiency led to hybrid approaches, such as Orso et al.'s two-phase RTS \cite{orso2004scaling} and Zhang's HyRTS, which combines method and class level analysis \cite{zhang2018hybrid}. Alongside dynamic RTS, static RTS techniques have been explored, though they face challenges with safety and practicality \cite{legunsen2017starts, legunsen2016extensive, shi2019reflection}.
	
	Current RTS research primarily focuses on the granularity of production code analysis, typically selecting test entities at class or method level. Building on the concept of checked coverage \cite{schuler2011assessing} and its correlation with test suite effectiveness \cite{zhang2015assertions}, our work innovates by selecting tests at the assertion level, treating each assertion and its slice (i.e., associated statements) as the smallest selection unit.
	
	Research on breaking down test cases for assertion-related studies includes Chen et al.'s assertion-aware test reduction \cite{chen2017assertions}, Vahabzadeh et al.'s fine-grained test case minimization \cite{vahabzadeh2018fine}, Fang et al.'s assertion fingerprints for identifying similar test cases \cite{fang2015identifying}, and Xuan et al.'s work on enhancing dynamic analysis and test purification \cite{xuan2016b, xuan2014test}. However, none of these specifically target RTS at the assertion level, making our approach a novel contribution to the field.

	\section{Conclusions}\label{conclusion}
	We proposed the first fine-grained regression test selection technique that works to select test entities at the assertion granularity level inside test methods. We used the proposed technique to implement a tool called \toolname capable of  effectively selecting change affected test entities to increase precision for selection and reduce test runtime. An empirical evaluation was performed on 11 subject systems. We compared \toolname with two state-of-the-art RTS tools. The results show that our technique  improves precision for all the subjects to select fewer tests by analyzing tests at the test assertion-level. \toolname  also outperforms existing tools in subjects with longer test runtime. In general, on average, it can reduce 63\% of the overall test time (RetestAll) among all subjects.
	For future work, we plan to apply fine-grained test analysis on other techniques such as test prioritization. We will also upgrade the implementation to support JUnit5 and extend the tool for other test domains including web and mobile tests that are known to have much longer running tests.
	
	\bibliographystyle{IEEEtran}
	\bibliography{main}

\begin{thebibliography}{10}
\providecommand{\url}[1]{#1}
\csname url@samestyle\endcsname
\providecommand{\newblock}{\relax}
\providecommand{\bibinfo}[2]{#2}
\providecommand{\BIBentrySTDinterwordspacing}{\spaceskip=0pt\relax}
\providecommand{\BIBentryALTinterwordstretchfactor}{4}
\providecommand{\BIBentryALTinterwordspacing}{\spaceskip=\fontdimen2\font plus
\BIBentryALTinterwordstretchfactor\fontdimen3\font minus
  \fontdimen4\font\relax}
\providecommand{\BIBforeignlanguage}[2]{{%
\expandafter\ifx\csname l@#1\endcsname\relax
\typeout{** WARNING: IEEEtran.bst: No hyphenation pattern has been}%
\typeout{** loaded for the language `#1'. Using the pattern for}%
\typeout{** the default language instead.}%
\else
\language=\csname l@#1\endcsname
\fi
#2}}
\providecommand{\BIBdecl}{\relax}
\BIBdecl

\bibitem{engstrom2010qualitative}
E.~Engstr{\"o}m and P.~Runeson, ``A qualitative survey of regression testing
  practices,'' in \emph{International Conference on Product Focused Software
  Process Improvement}.\hskip 1em plus 0.5em minus 0.4em\relax Springer, 2010,
  pp. 3--16.

\bibitem{elbaum2014techniques}
S.~Elbaum, G.~Rothermel, and J.~Penix, ``Techniques for improving regression
  testing in continuous integration development environments,'' in
  \emph{Proceedings of the 22nd ACM SIGSOFT International Symposium on
  Foundations of Software Engineering}.\hskip 1em plus 0.5em minus 0.4em\relax
  ACM, 2014, pp. 235--245.

\bibitem{memon2017taming}
A.~Memon, Z.~Gao, B.~Nguyen, S.~Dhanda, E.~Nickell, R.~Siemborski, and
  J.~Micco, ``Taming google-scale continuous testing,'' in \emph{2017 IEEE/ACM
  39th International Conference on Software Engineering: Software Engineering
  in Practice Track (ICSE-SEIP)}.\hskip 1em plus 0.5em minus 0.4em\relax IEEE,
  2017, pp. 233--242.

\bibitem{elbaum2000prioritizing}
S.~Elbaum, A.~G. Malishevsky, and G.~Rothermel, ``Prioritizing test cases for
  regression testing,'' in \emph{Proceedings of the 2000 ACM SIGSOFT
  international symposium on Software testing and analysis}.\hskip 1em plus
  0.5em minus 0.4em\relax ACM, 2000, pp. 102--112.

\bibitem{gligoric2014empirical}
M.~Gligoric, S.~Negara, O.~Legunsen, and D.~Marinov, ``An empirical evaluation
  and comparison of manual and automated test selection,'' in \emph{Proceedings
  of the 29th ACM/IEEE international conference on Automated software
  engineering}.\hskip 1em plus 0.5em minus 0.4em\relax ACM, 2014, pp. 361--372.

\bibitem{kiran2019comprehensive}
A.~Kiran, W.~H. Butt, M.~W. Anwar, F.~Azam, and B.~Maqbool, ``A comprehensive
  investigation of modern test suite optimization trends, tools and
  techniques,'' \emph{IEEE Access}, vol.~7, pp. 89\,093--89\,117, 2019.

\bibitem{yoo2012regression}
S.~Yoo and M.~Harman, ``Regression testing minimization, selection and
  prioritization: a survey,'' \emph{Software testing, verification and
  reliability}, vol.~22, no.~2, pp. 67--120, 2012.

\bibitem{kazmi2017effective}
R.~Kazmi, D.~N. Jawawi, R.~Mohamad, and I.~Ghani, ``Effective regression test
  case selection: A systematic literature review,'' \emph{ACM Computing Surveys
  (CSUR)}, vol.~50, no.~2, pp. 1--32, 2017.

\bibitem{gligoric2015practical}
M.~Gligoric, L.~Eloussi, and D.~Marinov, ``Practical regression test selection
  with dynamic file dependencies,'' in \emph{Proceedings of the 2015
  International Symposium on Software Testing and Analysis (ISSTA '15)}.\hskip
  1em plus 0.5em minus 0.4em\relax ACM, 2015, pp. 211--222.

\bibitem{ren2004chianti}
X.~Ren, F.~Shah, F.~Tip, B.~G. Ryder, and O.~Chesley, ``Chianti: a tool for
  change impact analysis of java programs,'' in \emph{Proceedings of the 19th
  annual ACM SIGPLAN conference on Object-oriented programming, systems,
  languages, and applications (OOPSLA04)}.\hskip 1em plus 0.5em minus
  0.4em\relax ACM, 2004, pp. 432--448.

\bibitem{zhang2011localizing}
L.~Zhang, M.~Kim, and S.~Khurshid, ``Localizing failure-inducing program edits
  based on spectrum information,'' in \emph{Proceedings of the 27th IEEE
  International Conference on Software Maintenance (ICSM '11)}.\hskip 1em plus
  0.5em minus 0.4em\relax IEEE Computer Society, 2011, pp. 23--32.

\bibitem{zhang2018hybrid}
L.~Zhang, ``Hybrid regression test selection,'' in \emph{Proceedings of the
  40th International Conference on Software Engineering (ICSE '18)}.\hskip 1em
  plus 0.5em minus 0.4em\relax IEEE, 2018, pp. 199--209.

\bibitem{commonsMath}
Math, ``The apache commons mathematics library,''
  \url{https://github.com/apache/commons-math/tree/master-old}, 2016, accessed:
  2022-11-10.

\bibitem{zhang2015assertions}
Y.~Zhang and A.~Mesbah, ``Assertions are strongly correlated with test suite
  effectiveness,'' in \emph{Proceedings of the 2015 10th Joint Meeting on
  Foundations of Software Engineering (ESEC/FSE'15)}.\hskip 1em plus 0.5em
  minus 0.4em\relax ACM, 2015, pp. 214--224.

\bibitem{weiser1984program}
M.~Weiser, ``Program slicing,'' \emph{IEEE Transactions on software
  engineering}, vol. SE-10, no.~4, pp. 352--357, 1984.

\bibitem{mohapatra2006overview}
D.~P. Mohapatra, R.~Mall, and R.~Kumar, ``An overview of slicing techniques for
  object-oriented programs,'' \emph{Informatica}, vol.~30, no.~2, pp. 253--277,
  2006.

\bibitem{hammer2004improved}
C.~Hammer and G.~Snelting, ``An improved slicer for java,'' in
  \emph{Proceedings of the 5th ACM SIGPLAN-SIGSOFT workshop on Program analysis
  for software tools and engineering}.\hskip 1em plus 0.5em minus 0.4em\relax
  ACM, 2004, pp. 17--22.

\bibitem{chen2001slicing}
Z.~Chen and B.~Xu, ``Slicing object-oriented java programs,'' \emph{ACM Sigplan
  Notices}, vol.~36, no.~4, pp. 33--40, 2001.

\bibitem{jdt}
JDT, ``Eclipse java development tools,'' \url{https://www.eclipse.org/jdt/},
  2022, accessed: 2022-11-10.

\bibitem{assertj}
AssertJ, ``fluent assertions java library,''
  \url{https://assertj.github.io/doc/}, 2022, accessed: 2022-11-10.

\bibitem{truth}
G.~Truth, ``Fluent assertions for java and android,'' \url{https://truth.dev/},
  2022, accessed: 2022-11-10.

\bibitem{repo}
Selertion, ``Fine-grained test selection at assertion-level,''
  \url{https://anonymous.4open.science/r/Selertion}, 2022.

\bibitem{ekstazi}
L.~E. Milos~Gligoric and D.~Marinov, ``Ekstazi: Lightweight test selection,''
  \url{http://www.ekstazi.org/maven.html}, 2014, accessed: 2023-10-24.

\bibitem{asterisk}
Asterisk-Java, ``The free java library for asterisk pbx integration,''
  \url{https://github.com/asterisk-java/asterisk-java}, 2020, accessed:
  2022-11-10.

\bibitem{net}
Net, ``Apache commons net,''
  \url{https://commons.apache.org/proper/commons-net/}, 2022, accessed:
  2023-08-01.

\bibitem{exec}
A.~C. Exec, ``Executing external processes from java,''
  \url{https://commons.apache.org/proper/commons-exec/}, 2014, accessed:
  2023-08-01.

\bibitem{tabula}
tabula java, ``Extract tables from pdf files,''
  \url{https://github.com/tabulapdf/tabula-java}, 2022, accessed: 2023-01-30.

\bibitem{otp}
OpenTripPlanner, ``An open source multi-modal trip planner,''
  \url{www.opentripplanner.org}, 2021, accessed: 2022-11-10.

\bibitem{streamlib}
Stream-lib, ``Stream summarizer and cardinality estimator,''
  \url{https://github.com/addthis/stream-lib/tree/v2.9.8}, 2019, accessed:
  2022-11-10.

\bibitem{tika}
A.~Tika, ``a content analysis toolkit,'' \url{https://tika.apache.org/}, 2020,
  accessed: 2022-11-10.

\bibitem{accumulo}
A.~Accumulo, ``A sorted, distributed key/value store that provides robust,
  scalable data storage and retrieval,'' \url{https://accumulo.apache.org/},
  2019, accessed: 2022-11-10.

\bibitem{pool}
A.~C. Pool, ``The apache commons object pooling library,''
  \url{https://github.com/apache/commons-pool}, 2019, accessed: 2022-11-10.

\bibitem{logic}
LogicNG, ``The next generation logic library,''
  \url{https://github.com/logic-ng/LogicNG}, 2019, accessed: 2023-01-31.

\bibitem{vahabzadeh2018fine}
A.~Vahabzadeh, A.~Stocco, and A.~Mesbah, ``Fine-grained test minimization,'' in
  \emph{Proceedings of the 40th International Conference on Software
  Engineering (ICSE '18)}.\hskip 1em plus 0.5em minus 0.4em\relax IEEE, 2018,
  pp. 210--221.

\bibitem{Mockito}
Mockito, ``Most popular mocking framework for unit tests written in java,''
  \url{https://github.com/mockito/mockito}, 2024, accessed: 2024-08-10.

\bibitem{junit5}
JUnit5, ``The 5th major version of the programmer-friendly testing framework
  for java and the jvm,'' \url{https://junit.org/junit5/}, 2022, accessed:
  2022-11-10.

\bibitem{major}
T.~M.~M. Framework, ``Easy and scalable mutation analysis for java,''
  \url{https://mutation-testing.org/}, 2022, accessed: 2023-08-01.

\bibitem{cloc}
Cloc, ``cloc counts blank lines, comment lines, and physical lines of source
  code in many programming languages,'' \url{https://github.com/AlDanial/cloc},
  2022, accessed: 2022-11-10.

\bibitem{surefire}
Surefire, ``Maven surefire plugin,''
  \url{https://maven.apache.org/surefire/maven-surefire-plugin/}, 2022,
  accessed: 2022-11-10.

\bibitem{harrold2001regression}
M.~J. Harrold, J.~A. Jones, T.~Li, D.~Liang, A.~Orso, M.~Pennings, S.~Sinha,
  S.~A. Spoon, and A.~Gujarathi, ``Regression test selection for java
  software,'' \emph{ACM Sigplan Notices}, vol.~36, no.~11, pp. 312--326, 2001.

\bibitem{legunsen2016extensive}
O.~Legunsen, F.~Hariri, A.~Shi, Y.~Lu, L.~Zhang, and D.~Marinov, ``An extensive
  study of static regression test selection in modern software evolution,'' in
  \emph{Proceedings of the 2016 24th ACM SIGSOFT International Symposium on
  Foundations of Software Engineering (FSE'16)}.\hskip 1em plus 0.5em minus
  0.4em\relax ACM, 2016, pp. 583--594.

\bibitem{orso2004scaling}
A.~Orso, N.~Shi, and M.~J. Harrold, ``Scaling regression testing to large
  software systems,'' \emph{ACM SIGSOFT Software Engineering Notes}, vol.~29,
  no.~6, pp. 241--251, 2004.

\bibitem{marascuilo1988statistical}
L.~A. Marascuilo and R.~C. Serlin, \emph{Statistical methods for the social and
  behavioral sciences.}\hskip 1em plus 0.5em minus 0.4em\relax WH Freeman/Times
  Books/Henry Holt \& Co, 1988.

\bibitem{zhu2019framework}
C.~Zhu, O.~Legunsen, A.~Shi, and M.~Gligoric, ``A framework for checking
  regression test selection tools,'' in \emph{2019 IEEE/ACM 41st International
  Conference on Software Engineering (ICSE)}.\hskip 1em plus 0.5em minus
  0.4em\relax IEEE, 2019, pp. 430--441.

\bibitem{just2014defects4j}
R.~Just, D.~Jalali, and M.~D. Ernst, ``Defects4j: A database of existing faults
  to enable controlled testing studies for java programs,'' in
  \emph{Proceedings of the 2014 international symposium on software testing and
  analysis}, 2014, pp. 437--440.

\bibitem{engstrom2008empirical}
E.~Engstr{\"o}m, M.~Skoglund, and P.~Runeson, ``Empirical evaluations of
  regression test selection techniques: a systematic review,'' in
  \emph{Proceedings of the Second ACM-IEEE international symposium on Empirical
  software engineering and measurement (ESEM '08)}.\hskip 1em plus 0.5em minus
  0.4em\relax ACM, 2008, pp. 22--31.

\bibitem{engstrom2010systematic}
E.~Engstr{\"o}m, P.~Runeson, and M.~Skoglund, ``A systematic review on
  regression test selection techniques,'' \emph{Information and Software
  Technology}, vol.~52, no.~1, pp. 14--30, 2010.

\bibitem{biswas2011regression}
S.~Biswas, R.~Mall, M.~Satpathy, and S.~Sukumaran, ``Regression test selection
  techniques: A survey,'' \emph{Informatica}, vol.~35, no.~3, pp. 289--321,
  2011.

\bibitem{rothermel1997safe}
G.~Rothermel and M.~J. Harrold, ``A safe, efficient regression test selection
  technique,'' \emph{ACM Transactions on Software Engineering and Methodology
  (TOSEM)}, vol.~6, no.~2, pp. 173--210, 1997.

\bibitem{vasic2017file}
M.~Vasic, Z.~Parvez, A.~Milicevic, and M.~Gligoric, ``File-level vs.
  module-level regression test selection for. net,'' in \emph{Proceedings of
  the 2017 11th Joint Meeting on Foundations of Software Engineering
  (ESEC/FSE'17)}.\hskip 1em plus 0.5em minus 0.4em\relax ACM, 2017, pp.
  848--853.

\bibitem{celik2017regression}
A.~Celik, M.~Vasic, A.~Milicevic, and M.~Gligoric, ``Regression test selection
  across jvm boundaries,'' in \emph{Proceedings of the 2017 11th Joint Meeting
  on Foundations of Software Engineering (ESEC/FSE'17)}.\hskip 1em plus 0.5em
  minus 0.4em\relax ACM, 2017, pp. 809--820.

\bibitem{legunsen2017starts}
O.~Legunsen, A.~Shi, and D.~Marinov, ``Starts: Static regression test
  selection,'' in \emph{Proceedings of the 32nd IEEE/ACM International
  Conference on Automated Software Engineering (ASE '17)}.\hskip 1em plus 0.5em
  minus 0.4em\relax IEEE Press, 2017, pp. 949--954.

\bibitem{shi2019reflection}
A.~Shi, M.~Hadzi-Tanovic, L.~Zhang, D.~Marinov, and O.~Legunsen,
  ``Reflection-aware static regression test selection,'' \emph{Proceedings of
  the ACM on Programming Languages}, vol.~3, no. OOPSLA, pp. 1--29, 2019.

\bibitem{schuler2011assessing}
D.~Schuler and A.~Zeller, ``Assessing oracle quality with checked coverage,''
  in \emph{Proceedings of the 2011 Fourth IEEE International Conference on
  Software Testing, Verification and Validation (ICST '11)}.\hskip 1em plus
  0.5em minus 0.4em\relax IEEE Computer Society, 2011, pp. 90--99.

\bibitem{chen2017assertions}
J.~Chen, Y.~Bai, D.~Hao, L.~Zhang, L.~Zhang, and B.~Xie, ``How do assertions
  impact coverage-based test-suite reduction?'' in \emph{Proceedings of the
  10th IEEE International Conference on Software Testing, Verification and
  Validation (ICST '17)}.\hskip 1em plus 0.5em minus 0.4em\relax IEEE, 2017,
  pp. 418--423.

\bibitem{fang2015identifying}
Z.~F. Fang and P.~Lam, ``Identifying test refactoring candidates with assertion
  fingerprints,'' in \emph{Proceedings of the Principles and Practices of
  Programming on The Java Platform (PPPJ '15)}.\hskip 1em plus 0.5em minus
  0.4em\relax ACM, 2015, pp. 125--137.

\bibitem{xuan2016b}
J.~Xuan, B.~Cornu, M.~Martinez, B.~Baudry, L.~Seinturier, and M.~Monperrus,
  ``B-refactoring: Automatic test code refactoring to improve dynamic
  analysis,'' \emph{Information and Software Technology}, vol.~76, pp. 65--80,
  2016.

\bibitem{xuan2014test}
J.~Xuan and M.~Monperrus, ``Test case purification for improving fault
  localization,'' in \emph{Proceedings of the 22nd ACM SIGSOFT International
  Symposium on Foundations of Software Engineering (SIGSOFT/FSE'14)}.\hskip 1em
  plus 0.5em minus 0.4em\relax ACM, 2014, pp. 52--63.

\end{thebibliography}
	
\end{document}